\documentclass[conference]{IEEEtran}
\IEEEoverridecommandlockouts

\usepackage{tcolorbox}
\usepackage{multirow} 
\usepackage{mathtools}
\usepackage{comment}
\usepackage{hyperref}
\usepackage{amsthm,amsmath}
\usepackage{amssymb}
\usepackage{booktabs} 
\usepackage{listings}
\usepackage{xcolor}
\usepackage{subcaption}

\definecolor{javared}{rgb}{0.6,0,0} %
\definecolor{javagreen}{rgb}{0.25,0.5,0.35} %
\definecolor{javapurple}{rgb}{0.5,0,0.35} %
\definecolor{javablue}{rgb}{0,0.2,0.6} %

\newtheorem{definition}{Definition}

\definecolor{LightGray}{gray}{0.9}

\tcbset{
    arc=0mm,
    left=1mm,
    right= 1mm,
    colback=LightGray,
    fontupper  = \slshape,
    leftrule=0mm,
    rightrule=0mm,
    toprule=0mm,
    bottomrule=1mm,
}

\AtBeginDocument{%
  }

\begin{document}

\title{Towards Human-interpretable Explanation in Code Clone Detection using LLM-based Post Hoc Explainer}

\author{\IEEEauthorblockN{
Teeradaj Racharak\IEEEauthorrefmark{1}\IEEEauthorrefmark{3}, Chaiyong Ragkhitwetsagul\IEEEauthorrefmark{2}\IEEEauthorrefmark{3}, Chayanee Junplong\IEEEauthorrefmark{2}\IEEEauthorrefmark{4}, Akara Supratak\IEEEauthorrefmark{2}}

\IEEEauthorrefmark{1}Advanced Institute of So-Go-Chi (Convergence Knowledge) Informatics, Tohoku University, Japan \\

\IEEEauthorrefmark{2}Faculty of Information and Communication Technology, Mahidol University, Thailand \\

racharak@tohoku.ac.jp, chaiyong.rag@mahidol.ac.th, nate.choco@gmail.com, akara.sup@mahidol.ac.th
\IEEEauthorblockN{\IEEEauthorrefmark{3}Equal contribution
\IEEEauthorrefmark{4}Corresponding author}
} 

\maketitle

\begin{abstract}
Recent studies highlight various machine learning (ML)-based techniques for code clone detection, which can be integrated into developer tools such as static code analysis. With the advancements brought by ML in code understanding, ML-based code clone detectors could accurately identify and classify cloned pairs, especially semantic clones, but often operate as black boxes, providing little insight into the
decision-making process. Post hoc explainers, on the other hand, aim to interpret and explain the predictions of these ML models after they are made, offering a way to understand the underlying
mechanisms driving the model’s decisions. However, current post hoc techniques require white-box access to the ML model or are computationally expensive, indicating a need for advanced post hoc explainers. In this paper, we propose a novel approach that leverages the in-context learning capabilities of large language models to elucidate the predictions made by the ML-based code clone detectors. We perform a study using ChatGPT-4 to explain the code clone results inferred by GraphCodeBERT. We found that our approach is promising as a post hoc explainer by giving the correct explanations up to 98\% and offering good explanations 95\% of the time. However, the explanations and the code line examples given by the LLM are useful in some cases. We also found that lowering the temperature to zero helps increase the accuracy of the explanation. Lastly, we list the insights that can lead to further improvements in future work. This study paves the way for future studies in using LLMs as a post hoc explainer for various software engineering tasks.
\end{abstract}

\maketitle

\section{Introduction}
Code clone detection is an activity of locating similar fragments of code based on their syntactic or semantic similarity~\cite{Rattan2013}.
Code clones can have both benefits and drawbacks.
On one hand, it leads to
the propagation of software defects, and increasing software maintenance~\cite{Roy2009}. On the other hand, it can be beneficial in software product lines~\cite{Kapser2008} and can be more stable than non-cloned code in some situations~\cite{Krinke2008}.
Code clone detection underpins many software engineering tasks, including software refactoring \cite{Balazinska2000}, code-to-code search~\cite{Luan2019}, software plagiarism~\cite{Prechelt2002}, software provenance~\cite{Davies2013}, and malware analysis~\cite{Farhadi2015}.
Previous studies have developed several techniques to automatically detect code clones using various representations of code, ranging from metrics~\cite{Saini2018}, text~\cite{Roy2008}, tokens~\cite{Sajnani2016}, n-grams~\cite{Ragkhitwetsagul2019}, trees~\cite{Jiang2007a}, graphs~\cite{Krinke2001}, and machine-learning embeddings for neural networks~\cite{Zhang2019}. The current trend aims towards detecting \textit{semantic clones}, i.e., clones with the same functionality but can be syntactically different.

On the other hand, with the progress of deep learning algorithms and the increase of computation power, the code representation methods for code semantics learning gradually develop from text and vocabulary to syntax and semantics \cite{lei2022deep,mostaeen2020machine,wang2020detecting,Guo2021}, and the corresponding models also tend to develop from sequence-based \cite{sajnani2016sourcerercc} to tree-based \cite{li2021secnn}, graph-based \cite{wang2020detecting}, and their combinations \cite{Guo2021}. 
Indeed, large BERT-based models such as CodeBERT and 
GraphCodeBERT is a pre-trained language model specifically designed for programming languages. CodeBERT~\cite{Feng2020} is trained on a large corpus of source code and natural language, enabling it to perform various downstream tasks such as code search, code summarization, and code translation. GraphCodeBERT
~\cite{Guo2021}, an extension of CodeBERT, incorporates both the textual and structural information of code by leveraging data flow graphs, enabling GraphCodeBERT to capture the semantics of code more effectively. As a result, it becomes useful for tasks that require an understanding of code structure, including code clone detection.

Both CodeBERT and GraphCodeBERT showed promising results on the clone detection tasks. When evaluated on the BigCloneBench dataset, CodeBERT achieved an F1-score of 0.965~\cite{Feng2020}. GraphCodeBERT advances upon this by specifically incorporating the inherent semantic structure of code, namely data flow, during pre-training. It treats code clone detection as a binary classification problem, where the probability of a true clone is determined by the dot product of the representations of the special token for two code fragments. GraphCodeBERT achieved superior clone detection performance to CodeBERT, RoBERTa (code), Transformer, and LSTM, with an F1-score of 0.971~\cite{Guo2021}. 

Nonetheless, although such pre-trained language models can accurately detect code clones, understanding the predictions of the models becomes increasingly intricate. Due to their inherent black-box nature, it is difficult to interpret their internal reasoning~\cite{xai4sebook}. 
This is especially important in the case of semantic clones because the similarity cannot be directly measured based on the amount of overlapping tokens, sequences, or their structure similarity using abstract syntax trees or control flow graphs.
Providing explanations on why the models classify a code pair as semantic clones will help aid the developers' understanding.

To this end, a plethora of post hoc explanation methods are proposed to provide explanations for these ML models’ predictions, highlighting the importance of each input feature
on the model’s prediction. Popular model-agnostic frameworks in 
explainable AI (XAI) include LIME (Local Interpretable Model-agnostic Explanations) \cite{ribeiro2016should}, SHAP (Shapley Additive Explanations) \cite{palacio2021xai}, 
and Counter Factual (CF) frameworks such as DiCE (Diverse Counterfactual Explanations) \cite{chou2022counterfactuals}. 
These works utilize different local 
neighborhoods to realize the cause-and-effect relationships within ML processes. 
Unfortunately, existing post hoc approaches may require a white-box
access to the ML model or are computationally expensive, indicating
a need for advanced post hoc explainers. 
For example, TreeSHAP requires access to the internal structure of the tree model (i.e., its nodes, splits, and paths) — i.e., you need white-box access to its structure.
Inspired by the local neighborhood approximations in XAI, \textbf{this paper fills the gap by proposing a novel approach that leverages In-context Learning (ICL) \cite{mann2020language} capabilities of large language models (LLMs) to elucidate the predictions made by the ML-based code clone detectors.} Indeed, we primarily focus on ``local'' post hoc explainers in the code clone detection setting, i.e., methods explaining important code features in cloned (or non-cloned) pairs predicted by a code clone detector. 

To help us pave the way for an advanced ICL-based 
post hoc explainer, we set up key research questions (RQs) as 
follows: 

\textit{\textbf{RQ1:} To what extent can LLMs explain the behavior of code clone detectors?} This paper investigates this question by using an LLM to explain the clone detection result by GraphCodeBERT. 

\begin{itemize}
    \item \textit{\textbf{RQ1.1:} Given a clone prediction from a blackbox detector, how well does an LLM realize a relevant explanation?} This sub RQ focuses on the explanation generated by an LLM for a positive prediction, i.e., clone pairs.
    \item \textit{\textbf{RQ1.2:} Given a non-clone prediction from a blackbox detector, how well does an LLM realize a relevant explanation?} This sub RQ focuses on the explanation generated by an LLM for a negative prediction, i.e., non-clone pairs.
\end{itemize}
    
\textit{\textbf{RQ2:} What is the effective number of local neighborhood examples?} In this RQ, we are interested in comparing the number of local neighborhood examples, described in terms of ``few-shot'', for explaining the internal process of code clone detectors.

\textit{\textbf{RQ3: } What are the effects of the LLM temperature on generating the explanations?} Since the temperature dictates how creative an LLM is in generating its response, in this RQ, we empirically investigate the effects of changing the LLM's temperature on its generated code clone explanations.

\textit{\textbf{RQ4: } How are the sampled code lines selected for the explanations?} We instructed the LLM to generate explanations along with selecting ten code lines from the given code pair that influence the classification of clone/non-clone. This RQ investigates the code lines that are selected to include in the explanations.

LLMs have emerged as powerful tools that are effective at a wide variety of tasks. However, their potential to explain the behavior of other complex predictive models remains relatively unexplored, especially in software engineering contexts. 
As part of our study, we clarify and systematize the ICL approach and study the few-shot setting in detail. 
We perform a study using ChatGPT-4 to explain the code clone results inferred by GraphCodeBERT. 

\section{Background}

Our work lies in the local neighborhood concept and the ICL capabilities of LLMs. Their basics are briefly provided as follows.

\subsection{Explainable Artificial Intelligence (XAI)}
\label{subsection:xai}
Explainable Artificial Intelligence refers to the capability of AI systems to provide clear and comprehensible explanations for their actions and judgments. The primary objective is to enhance human comprehension of these systems by clarifying the fundamental mechanisms governing their decision-making processes.
However, rather than successfully meeting the demands of the intended users, many attempts to increase explainability often result in explanations that are primarily fitting to the AI researchers themselves. This imposes the responsibility of developing convincing justifications for intricate decision models on AI specialists who possess a thorough comprehension of these models \cite{miller2017explainable}. The most often used methodologies for XAI are SHAP~\cite{shap} and LIME~\cite{Lime}.

\subsection{Local Neighborhoods} 
\label{subsection:local-neighborhoods}

The two most commonly used post hoc techniques in XAI are feature importance (such as SHAP and LIME) and counterfactuals (such as DiCE), which employ the concept of local neighborhoods to compute the sufficient and necessary scores of the features involved in the ML's inference mechanism. Indeed, post hoc explanation frameworks generate local explanations using neighborhoods around the input sample. This local neighborhood can be 
defined by various criteria, such as, 
marginal or conditional distributions, as it represents the distribution following specific perturbations after the samples are drawn from a predetermined distribution \cite{pawar2024impact}. 

For instance, the neighborhood in SHAP is generated after perturbing an input and weighting the local neighborhood samples using the kernel-weighted conditional probability based on a given distribution that is adjusted for interventional effects due to feature perturbations. Based on the kernel, higher values of a 
neighbourhood are assigned to samples with very small or very large replacement of features. On the other hand, LIME generates a local 
neighborhood around an input in question and trains different interpretable linear models on the local neighborhood samples weighted using a distance metric. A local neighborhood in LIME is computed using training samples and perturbed samples. 

\subsection{In-context Learning (ICL)}

Recent results in natural language processing (NLP) have shown that the 
parameterized knowledge within LLMs can be zero-shot transferred 
to perform standard NLP tasks without the need for fine-tuning on 
a training set. LLMs offer the exciting prospect of in-context few-shot learning via prompting. Instead of fine-tuning a separate model checkpoint for each new task, one can simply `prompt' the model with a few input–output exemplars demonstrating the task. 
Remarkably, this has been successful for a range of simple question-answering tasks \cite{brown2020language} to tasks that require emulating the thought processes of human reasoners, such as arithmetic, commonsense, and symbolic reasoning \cite{wei2022chain}. 
State-of-the-art LLMs such as OpenAI's ChatGPT \cite{GPT}, Claude \cite{Claude-2}, and Gemini \cite{Gemini}, are extensive used across many applications, such as commonsense reasoning \cite{Wei2022ChainOT}, text translation \cite{Hendy2023HowGA}, language understanding \cite{brown2020language}, coding tasks \cite{Bubeck2023SparksOA}, and LLMs as explanation methods \cite{kroeger2024large}. 

\section{LLMs as Code Clone Explainers}

This section describes our proposed knowledge-based local neighborhood concept and how to draw samples from the neighborhoods to create prompts for code clone explanation.

\subsection{Knowledge-based Local Neighbourhood (KLN)}
\label{sec:nbh}

It is worth noting that investigating the potential of LLMs to explain the behavior of other complex predictive models remains relatively 
unexplored. An exceptional work that is close to our work is Kroeger et al.~\cite{kroeger2024large}, which studies two kinds of ICL-based prompting 
strategies, namely perturbation-based and explanation-based ICL. 
In this work, we aim to demonstrate that in-content learning efficiency can be 
improved by utilizing domain knowledge to guide the learning process in LLMs. This perspective is not explored yet, though there exist similar 
studies in other areas of machine learning. For example, Vu et al.~
\cite{vu2020progressive} demonstrated how to inject knowledge during the 
training process for enhancing the accuracy. We give a formal definition of our proposed \emph{knowledge-based local neighborhood (KLN)} as follows. 

\begin{definition}
\label{def:knowledge-neighbor}
    Let $\mathcal{D} \coloneqq \{ (x_i, c_i) \mid i \in \mathbb{N} \}$ be a dataset where $x_i$ denotes $i$-th source code and $c_i$ represents 
    its corresponding category. For any $i, j \in \mathbb{N}$, 
    we say $x_i$ and $x_j$ are $c$-\emph{degree neighborhoods} of each other if the knowledge-based distance between them (denoted by $d_k (c_i, c_j)$) is equal to $c \in \mathbb{R}$. 
\end{definition}

Note that the KLN uses the domain 
knowledge available in the dataset for categorizing instances in the dataset. 
This definition is very general in that it does not provide any concrete notion of the distance function $d_k$. For instance, one may acquire a taxonomy (or an ontology) of such labels and employ semantic similarity between two categories 
in order to sort out which individual $x_i$ is a ``closer'' neighbor to individual $x_j$. Another idea is to 
calculate the similarity between two abstract syntax
trees (ASTs), or code embeddings, of the corresponding source codes and define criteria
for different degrees. 

In the following, we show our application of Definition \ref{def:knowledge-neighbor} to the code clone detection setting. For that, we treat this 
problem as a binary classification so that any code pair is mapped to 
1 if they are cloned and 0 otherwise. We construct a new dataset $\mathcal{D}^\prime$ from dataset $\mathcal{D}$ for explaining code 
clone as follows:  
\[
\mathcal{D}^\prime \coloneqq \{ \big( (x_i, x_j), y \big) \text{ for any } x_i, x_j \in \mathcal{D} \text{ and } y \in \{ 0, 1 \}  \}
\]
\noindent
where $1$ represents cloned one and $0$ otherwise. These labels are normally 
provided in the existing code clone datasets. For example, 
BigCloneBench provided four clone types, namely Type I, II, III, and IV (semantic clones). We further explain how such labels 
are manipulated in this work in Section \ref{subsubsection:dataset}.

To apply Definition \ref{def:knowledge-neighbor}, we consider three degrees of ordering for the local neighborhoods in our 
setting. For any two code pairs $(x_i, x_j)$ and $(x_m, x_n)$ in 
$\mathcal{D}^\prime$, we say that
\begin{enumerate}
    \item $(x_i, x_j)$ has a \emph{high}-degree neighborhood with 
    $(x_m, x_n)$ if their $c_i, c_j, c_m, c_n$ represent the same category; 

    \item $(x_i, x_j)$ has a \emph{medium}-degree neighborhood with 
    $(x_m, x_n)$ if either $c_m$ or $c_n$ represents the same category with $c_i, c_j$; 

    \item $(x_i, x_j)$ has a \emph{null}-degree neighborhood with 
    $(x_m, x_n)$ if neither $c_m$ nor $c_n$ represents the same category with $c_i, c_j$. %
\end{enumerate}
\noindent
Since many code clone benchmarks provide the clone classes (i.e., groups of clones), 
we can treat each clone class as a ``category'' in our setting. It is feasible to consider other kinds of dynamic category-based distance metrics. For example, using AST trees or code embeddings of the code pair for calculating the proximity as their degree of neighborhood. 
We leave this task of investigating
distance functions for our future study.

\subsection{From KLN samples to Few Shot}

Intuitively, our framework is motivated by Ribeiro et al. \cite{Lime} where one could train a highly accurate interpretable function from a complex non-linear one. Concretely, given a target (non-linear) prediction 
model $f$ and a test code pair $(x_i, x_j)$ to be explained, we aim at obtaining an an interpretable decision boundary $f^\prime$ approximating from 
$f$ that could yield likely the same outcomes 
$f(x_i, x_j) \approx f^\prime (x_i, x_j)$. We 
follow a similar manner as training such an approximately interpretable function as in \cite{Lime} by randomly sampling a sufficient number of code pairs from the neighborhoods $\mathcal{N}_{(x_i, x_j)}$ of $(x_i, x_j)$ 
and get the corresponding model predictions of $\mathcal{N}_{(x_i, x_j)}$ by using $f$. We aim 
to validate whether LLM could find $f^\prime$ from the complex non-linear decision boundary $f$ in this study. In this paper, we call the sampled code pairs from the neighborhoods the \textit{KLN samples}.

\begin{figure}[tb]
    \centering
    \includegraphics[width=\columnwidth]{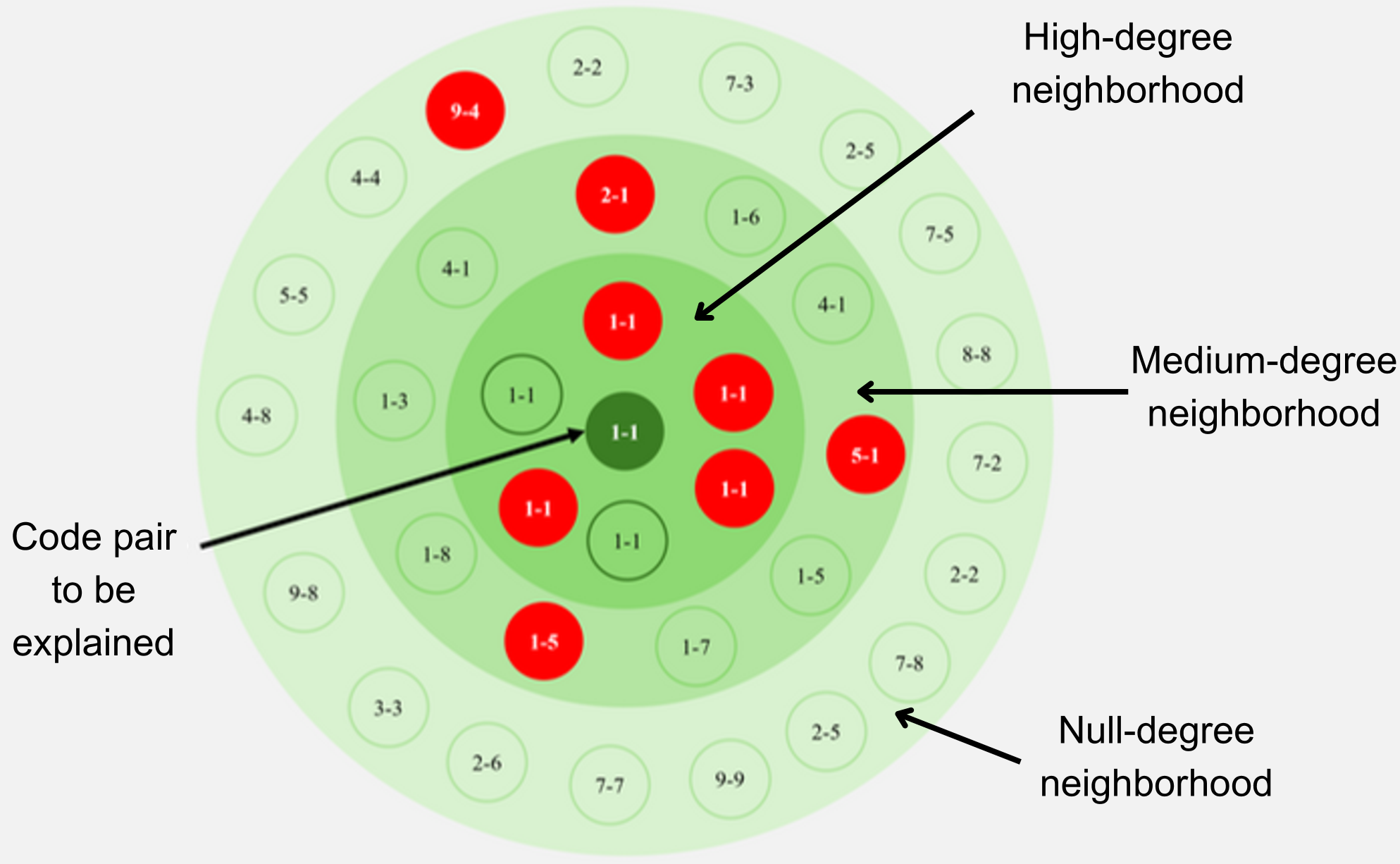}
    \caption{KLN, where the value in the circles denotes a corresponding category of code pairs (e.g., 1--1 is a clone pair)}
    \label{fig:code-clone}
\end{figure}

Figure~\ref{fig:code-clone} explains the above KLN concept adopted into our code-clone explanation context.
Each circle represents a code pair where a prediction model $f$ could classify it as either 
a clone or a non-clone. The actual ground truth is represented by the value in each circle. For example, 1--1 and 2-2 are clone pairs, while 1-2 and 2-3 are non-clone pairs. In this example, the test code pair $(x_i, x_j)$ that needs to be explained is the green circle with the value 1--1, i.e., the clone pair from the same class 1. 
We select the KLN samples from the code pairs surrounding $(x_i, x_j)$ as illustrated in 
the figure. The sampled code pairs are highlighted in red. This scopes down the number of examples that are given to the explainer and simplifies $f$ for an accurate explanation of $(x_i, x_j)$. 
To alleviate the computational problems of post hoc explainers, we explore using a small number 
of KLN samples, i.e., 4 and 8 samples in our experiment, from $\mathcal{N}_{(x_i, x_j)}$ 
in our prompting templates for the LLM. 

To avoid an over-fitting-like issue from ICL within the LLM, 
we balance the KLN samples with respect to their classes, given a test code pair. 
Moreover, to guide the generation of the explanation, we put the highest emphasis on the high-degree KLN samples (cloned), followed by the medium-degree (non-cloned) and null-degree (non-cloned) KLN samples, with the approximate ratio of 50\%, 35\%, and 15\% of the number of samples, respectively. 
Thus, for the neighborhoods $\mathcal{N}_{(x_i, x_j)}$ with size 4, 
we randomly sample 2 high-degree neighborhoods, 1 medium-degree neighborhood, and 
1 null-degree neighborhood. Analogously, for the neighborhoods $\mathcal{N}_{(x_i, x_j)}$ with size 8, we randomly sample 4 high-degree 
neighborhoods, 3 medium-degree neighborhoods, and 1 null-degree neighborhood 
for the neighborhoods $\mathcal{N}_{(x_i, x_j)}$. 
These equal numbers of cloned pairs and non-cloned pairs are used to guide a few-shot learning fairly. 

Given the neighborhoods $\mathcal{N}_{(x_i, x_j)}$, 
their corresponding prediction output from $f$, and 
a test code pair $(x_i, x_j)$, we can now generate 
a prompt to use with the LLM as a post hoc code clone 
explainer. Figure \ref{fig:prompt} shows its overall prompt structure. There are four distinct parts in our template. 
(1) \emph{Context}: We provide relevant information about 
the black-box model to be explained. (2) \emph{Dataset}: 
We provide a list of inputs and their respective models, output from KLN samples ($\mathcal{N}_{(x_i, x_j)}$ and $f$). We 
also provide the confidence score, which is derived from the softmax function of the model to assist the LLM 
in understanding the target predictive model $f$. 
After that, we append the test code pair to be explained (highlighted in red). 
(3) \emph{Question}: we give a specific command for the 
LLM to approximate an interpretable decision boundary 
$f^\prime$. (4) \emph{Instruction}: we give the directives 
for explanation's generation. 

It is obvious that our framework, which exploits ICL to approximate an interpretable decision boundary 
$f^\prime$, does not require to fully access the black-box model $f$. Therefore, it fully functions 
as an (absolute) post-hoc explainer.

\subsection{Prompt Engineering}
\label{subsection:prompts-template}

Previous post hoc explanations have recommended using a huge number of KLN samples \cite{smilkov2017smoothgrad}. However, it is not practical to include all samples from \(N_x\) in the prompt for an LLM due to limitations on the maximum context length and the resulting decrease in performance when too much information is provided \cite{liu2023lost}. 
Moreover, we utilize prior studies on LLMs \cite{kroeger2024large} in a perturbation-based ICL prompting approach that effectively utilizes LLMs' ability to efficiently explain the behavior of machine learning predictive models. We prompt LLMs to analyze and select the most relevant five lines from each code snippet, i.e., \textit{code line examples}, for explaining why the model makes such a prediction. 
Inspired by the local neighborhood approximation techniques in XAI, if there are enough disturbances in \(N_x\), it is anticipated that LLM can effectively represent the behavior of a simplified and interpretable version of the black box model. 

\section{Methodology}
\label{section:prelim-study}

We evaluate the effectiveness of our LLM-based post hoc explainer by performing an empirical study. The methodology of the study is as explained below.

\subsection{Tools, Prompt Template, and Dataset}
\subsubsection{Code Clone Detection Model}
We use GraphCodeBERT~\cite{Guo2021} as a code clone detector. 
GraphCodeBERT is trained on the code's data flow information.
The result of code clone detection shows that GraphCodeBERT outperforms CodeBERT and the RoBERTa (trained on code) models and five state-of-the-art clone detectors on the BigCloneBench dataset~\cite{Svajlenko2022} with a F1-score of 0.971. Initially, we also tried to apply SHAP to generate an explanation for GraphCodeBERT, but without success because it did not support the specific structure in the GraphCodeBERT model.

We adopted the code examples provided in the tool's GitHub repository\footnote{https://github.com/microsoft/CodeBERT/tree/master/GraphCodeBERT} and performed the clone detection experiment on Google Colab. The GraphCodeBERT was used as-is to perform the classification of the given code pairs as a non-linear blackbox prediction model.

\subsubsection{Dataset}
\label{subsubsection:dataset}

As we focus on explaining semantic clone detection, we use the Google Code Jam (GCJ)\footnote{https://code.google. com/codejam} dataset from the study by Zhao and Huang~\cite{Zhao2018}. 
GCJ is an annual online programming competition, in which the results are made publicly available.
A set of programming questions is given, and the participants submit their answers (one answer per file per question). Thus, the GCJ data establishes a semantic clone ground truth. 
The answers to the same questions are considered clones of each other, while the answers to different questions are considered non-clones of each other. 
The GCJ dataset is considered a challenging clone benchmark used in many previous studies~\cite{Guo2021,Zhao2018,Krinke2021,Choi2023,Su2016a,Su2016b}.

\subsubsection{Prompt Template}
\begin{figure}[tb]
\centering
    \includegraphics[width=\columnwidth]{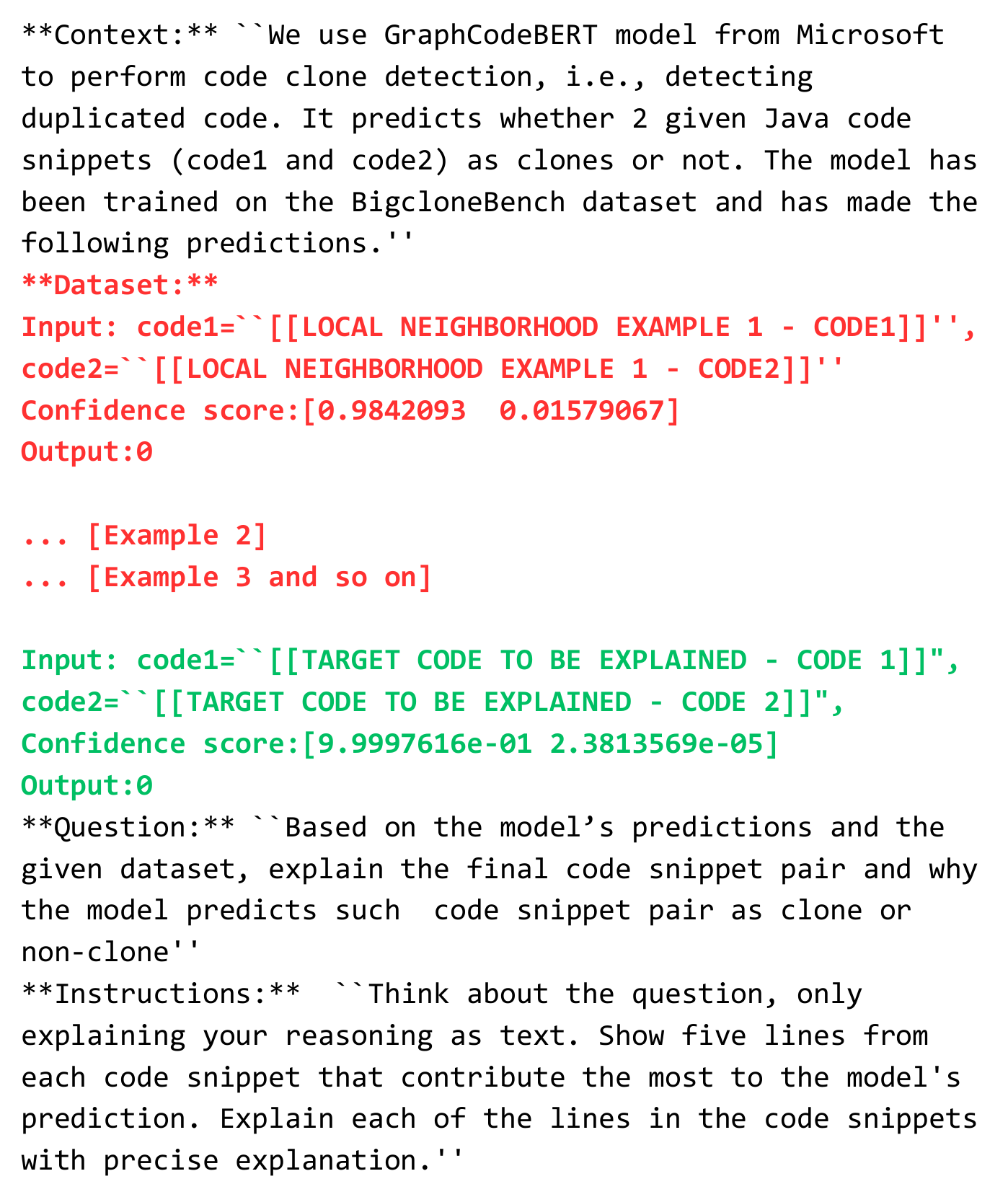}
    \caption{A sample prompt generated with the KLN strategy}
\label{fig:prompt}
\end{figure}

As previously discussed, Figure~\ref{fig:prompt} shows our prompt template consisting of four components: Context, Dataset, Question, and Instruction. Our Context explains that the code clone detection model is GraphCodeBERT and its relevant information. The Dataset part contains the KLN samples and the target code pair to be explained. The Question specifies that the LLM must explain why the target code pair is predicted as clone/non-clone. The Instruction tells the LLM to select the code line examples (i.e., \textit{``. . . Show five lines from each code snippet that contribute the most to the model's prediction.''}). Our intuition is that having the five code line examples in the explanation may help the developers to better understand why a pair is predicted as a clone/non-clone. 

\subsection{Experimental Setup}
\begin{figure*}
    \centering
    \includegraphics[width=\textwidth]{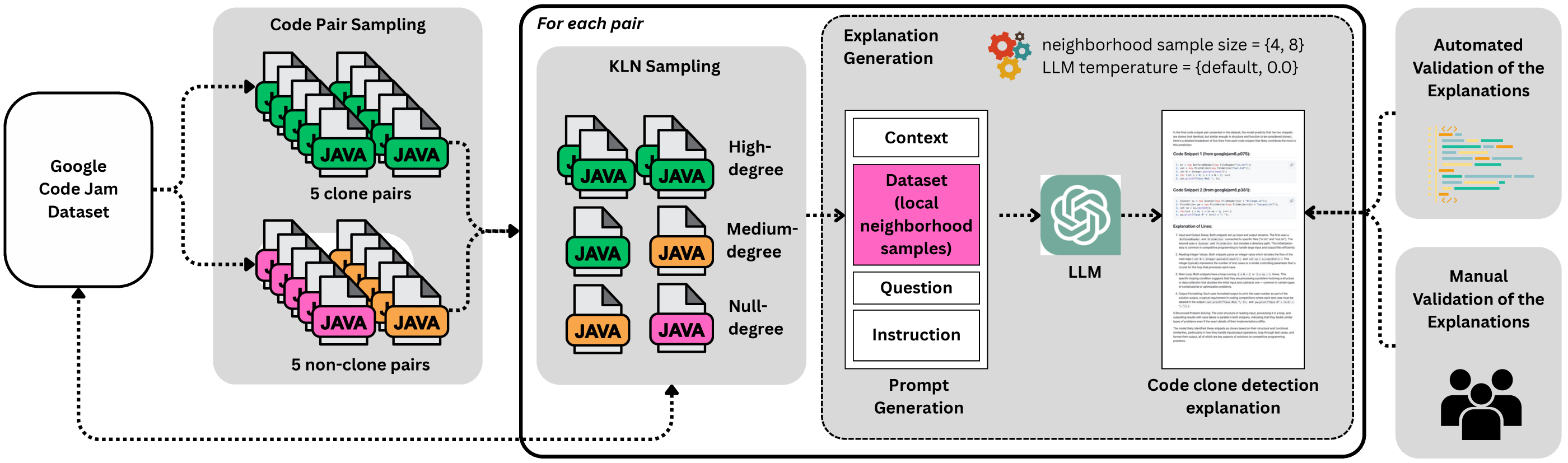}
    \caption{Experimental setup of this study}
    \label{fig:experiment}
\end{figure*}

Our experimental setup is shown in Figure~\ref{fig:experiment}. It is divided into four main phases: code pair sampling, KLN sampling, explanation generation, and validation. We explain each phase in detail below.

\subsubsection{Code Pair Sampling}
Our dataset contains 12 questions with multiple answers, and we focused only on the Java answers. 
It comprises 1,665 Java files, including 274,959 clone pairs and 1,110,321 non-clone pairs.
From this dataset, we randomly sampled five clone pairs and five non-clone pairs to be the test code pairs, which were explained by the LLM. All five clone pairs came from different questions.

\subsubsection{KLN Sampling}
\label{sec:lns}
To perform a KLN sampling, we search the Google Code Jam dataset, based on the target code pair to be explained, using our KLN strategy with the three degrees of ordering (see Section~\ref{sec:nbh} and \ref{sec:lns}). 

If the target code pair was a clone pair, for the high-degree neighborhood, we randomly picked code pairs of the same question as the target code pair, i.e., clone pairs (as shown by the green color code snippets in Figure~\ref{fig:experiment}). For the medium-degree neighborhood, we randomly picked one code pair of the same question as the target code pair, and one code pair from another question, i.e., a non-clone pair containing one code snippet from the same question as the target code pair (as shown by one green color code snippet and one yellow color code snippet in Figure~\ref{fig:experiment}). For the null-degree neighborhood, we randomly picked one code snippet from one question and another code snippet from another question. Both questions must not be the same as the question of the target code pair, i.e., a non-clone pair with different questions (as shown by one yellow color code snippet and one pink color code snippet in Figure~\ref{fig:experiment}). 
In contrast, if the target code pair was a non-clone pair (i.e., they were from different questions), we performed the inverse. We started with the null-degree neighborhood and randomly picked one clone pair from a different question than the target code pair. For the medium-degree neighborhood, one code snippet was from the clone pair in the null-degree neighborhood, and the other one was from a different question. For the high-degree neighborhood, the two code snippets were randomly selected from the same respective question of the target code pair.

To further illustrate our KLN sampling, three code examples extracted from the Google Code Jam dataset~\cite{Zhao2018} are shown in Figure~\ref{fig:gcj_examples} as a simple example. The first and the second code snippets ($C_1$ and $C_2$) are answers to the same programming question, i.e., question 3 (Brattleship). They share the same functionality and are treated as a clone pair. We can observe that the two code snippets also share some syntactic similarity, such as the \texttt{for} loop based on the user-given value and the three statements of \texttt{nextInt()} calls. In contrast, the third code snippet ($C_3$) is an answer to question 12 (TeachingAssistant). It does not share the same functionality as $C_1$ and $C_2$. Thus, ($C_1$, $C_3$) or ($C_2$, $C_3$) is a non-clone pair. In the case that we have a test code pair ($C_i$, $C_j$) from question 3 and we want to locate KLN samples, we could potentially pick ($C_1$, $C_2$) as a high-degree neighborhood sample, ($C_1$, $C_3$) as a medium-degree neighborhood sample, and another code pair that comes from other questions except question 3 as the null-degree neighborhood sample.

\begin{figure*}[tb]
    \centering
    \begin{subfigure}[t]{0.32\textwidth}
\begin{lstlisting}
...
public class Brattleship {
  public static final String FILENAME = "A-large";

  public static void main(String[] args) throws IOException {
    BufferedReader in = new BufferedReader(new FileReader("src/" + FILENAME
      + ".in"));
    BufferedWriter out = new BufferedWriter(new FileWriter("src/"
      + FILENAME + ".out"));
    Scanner sc = new Scanner(in);
    int t = sc.nextInt();
    for (int tt = 1; tt <= t; tt++) {
      int r = sc.nextInt();
      int c = sc.nextInt();
      int w = sc.nextInt();
      int ret = (r-1)*(c/w);
      if ((c/w)*w==c) {
        ret += c/w-1+w;
      } else {
        ret += c/w+w;
      }
      out.write("Case #"+tt+": "+ret+"\n");
    }
    in.close();
    out.close();
  }
}
\end{lstlisting}
    \caption{$C_1$: Question 3 Brattleship (p033)}
    \label{fig:gcj_example_3-1}
    \end{subfigure}
\begin{subfigure}[t]{0.32\textwidth}
\begin{lstlisting}
...
public class A {
  public static void main(String[] args) {
    A a = new A();
    Scanner in = new Scanner(System.in);
    int cases = Integer.parseInt(in.nextLine());
    for (int a1=0; a1<cases; a1++) {
      int r = in.nextInt();
      int c = in.nextInt();
      int w = in.nextInt();
      if (a1!=cases-1) {
        in.nextLine();
      }
      int base = c/w;
      base *= r;
      if (c%
        base += (w-1);
      }
      else {
        base += w;
      }
      System.out.println("Case #"+(a1+1)+": "+base);
    }
  }
}
\end{lstlisting}
    \caption{$C_2$: Question 3 Brattleship (p039)}
    \label{fig:gcj_example_3-2}
    \end{subfigure}
    \begin{subfigure}[t]{0.32\textwidth}
\begin{lstlisting}
...
public class A {
  public static void main(String[] args) throws Exception {
    Scanner in = new Scanner(new File("AL.in"));
    PrintWriter out = new PrintWriter("AL.out");
    int tc = in.nextInt();
    for (int cc = 1; cc <= tc; cc++) {
      String s = in.next();
      int gain = 0;
      while (true) {
        Stack<Character> stack = new Stack<>();
        for (int i = 0; i < s.length(); i++) {
          if (!stack.isEmpty()&&stack.peek()==s.charAt(i)) {
            stack.pop();
          } else { stack.push(s.charAt(i)); }
        }
        StringBuilder next=new StringBuilder();
        for (Character c:stack) { next.append(c); }
          String nx = next.toString();
          int value = (s.length()-nx.length())/2*10;
          s = nx;
          if (value==0) break;
          gain+=value;
        }
        gain+=s.length()/2*5;
        System.out.printf("Case #%
        out.printf("Case #%
    }
    out.close();
  }
}
\end{lstlisting}
    \caption{$C_3$: Question 12 TeachingAssistant (p022)}
    \label{fig:gcj_example_3-3}
    \end{subfigure}
    \caption{Code snippets from Google Code Jam. ($C_1$, $C_2$) is a clone pair, while ($C_1$, $C_3$) is not.}
    \label{fig:gcj_examples}
\end{figure*}

As previously mentioned, in this study, we investigated two different KLN sizes: size-4 and size-8. 
For size-4, the sampling is done by randomly selecting 2 examples from the high-degree neighborhood, 1 from the medium-degree neighborhood, and 1 from the null-degree neighborhood, respectively.
For size-8, the sampling is done similarly by randomly selecting 4, 3, and 1 examples (we add the additional 4 KLN samples into the KLN samples of size 4).

\subsubsection{Explanation Generation}
\begin{figure}[tb]
    \centering
    \includegraphics[width=\columnwidth]{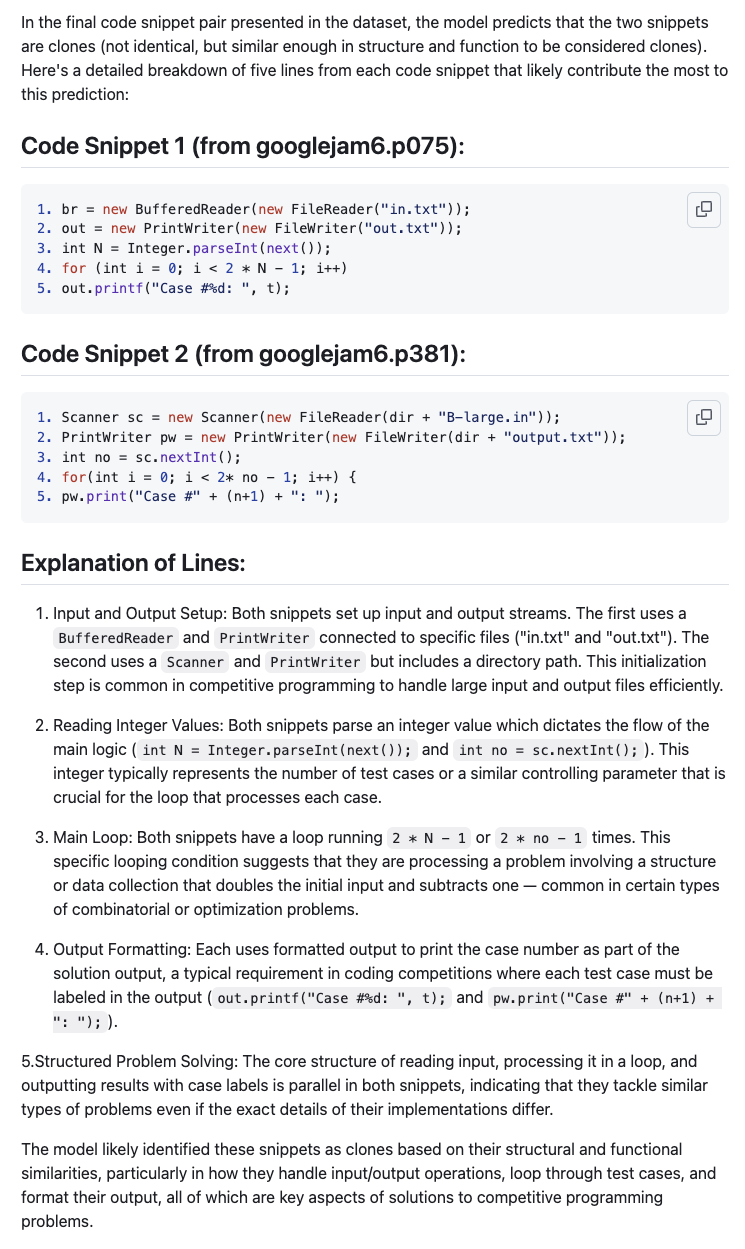}
    \caption{An explanation of a clone pair generated by LLM}
    \label{fig:explanation}
\end{figure}

After obtaining the KLN samples, we constructed the prompt based on the prompt template. First, the code of the test code pair to be explained and the clone detection result from GraphCodeBERT were added to the prompt. Then, the code snippets of all the KLN samples were added.
Lastly, we fed the filled prompt into OpenAI's GPT-4 model. We then parsed its responses to discover the explanation. The explanation was saved to a text file in the Markdown format. 
An example of the explanation generated by GPT-4 is shown in Figure~\ref{fig:explanation}.

\subsubsection{Automated Validation of LLM Explanations}
The automated validation of LLM explanations is used to check two aspects of the explanations: the correctness of the explanation (i.e., clone vs.~non-clones), and the correct selection of the sampled code lines (i.e., they exist in the original code or not). It is considered a sanity check. A more thorough and qualitative evaluation of the explanations is performed using a manual validation (see Section~\ref{sec:manual_validation}).

For each test code pair to be explained, we varied the settings as follows. First, we chose two sizes of the local KLN samples: 4 and 8. Second, we adjusted the temperature of the LLM. We tried the default temperature and the zero temperature. Lastly, each pair was run five times to account for the nondeterminism of LLMs, and the KLN samples were randomly sampled for every run. This process yielded a total of 200 responses (10 code pairs $\times$ 2 local neighborhood sizes $\times$ 2 temperatures $\times$ 5 runs), with 100 responses generated from KLN samples of size 4 and another 100 from size 8. 

We perform the automated validation of the explanations by using a Python script to automate the process of local neighborhood sampling, prompt generation, and explanation generation. We used OpenAI ChatGPT's API to connect to the LLM. 
Another script then reads the generated explanations and evaluates whether they accurately correspond to the model’s prediction. This evaluation will particularly focus on the words \textit{``clone''} or \textit{``non-clone''} in the explanation. Then, the script verified that the five selected code lines from each code snippet in the test pair are present within the original code snippets using string matching. The script also recorded the location where the selected code lines appear in the original code.

\subsubsection{Manual Validation of LLM Explanations}
\label{sec:manual_validation}
We also performed manual validations to determine the quality of the explanations. This process is carried out by having one author as the main validator and another author as the second validator. Similar to the automated validation, we randomly selected 10 different code pairs (5 clone and 5 non-clone pairs), and generated the explanations using the default temperature. We created two sets of explanations: size-4 and size-8. For the manual validation, the explanation was generated only once.

We defined a set of manual validation steps so that the validators performed the validation in the same way. The steps are as follows. 

\textbf{1.~Set the Context}: The validators studied the GCJ questions associated with the test code pairs. The validators then checked the category, i.e., whether the test code pair was clones or non-clones, based on the ground truth and the model's prediction. 
    
\textbf{2.~Check the Correctness of the Explanation}: The validators determined whether the explanation is correct or incorrect according to the model's prediction. If a test code pair is predicted as clones, the explanation should say the same, and vice versa. The validators also tried to understand the rationale of such a decision given by the LLM 
to verify if the explanation aligns with the model's prediction, specifically by providing a rationale for why a given pair is categorized as a clone or non-clone.

\textbf{3.~Assess the Quality of the Explanation:} The validators carefully read the explanations and classified each of them into one of the two classes (see Table \ref{tab:classification}): (1) Good Explanation and (2) Bad explanation (i.e., no example, irrelevant, or wrong explanation). The validators also noted the cases where the five chosen code line examples are not useful.

We followed the steps above and validated 20 explanations. In case of disagreements, the third validator was employed. If there was still no consensus, the three validators discussed until an agreement was reached. %

\subsection{Results}
\label{sec:result}
Next, we discuss the experimental results and answer the research questions (RQ1--RQ3).

\begin{table}[tb]
\caption{Explanation accuracy based on the automated validation}
  \label{tab:percentage}
  \centering
\begin{tabular}{@{}lrrrr@{}}
\toprule
\multicolumn{1}{c}{\multirow{2}{*}{Result}}           & \multicolumn{2}{c}{Temp=default} & \multicolumn{2}{c}{Temp=0} \\ \cmidrule(l){2-5} 
\multicolumn{1}{c}{} & Size-4 & Size-8 & Size-4 & Size-8 \\ \midrule
Explanation correctness & 90\% & 78\% & \textbf{98\%} & 78\% \\

~~~-~Clone pairs & 92\% & 96\% & \textbf{100\%} & 92\% \\
~~~-~Non-clone pairs & 88\% & 60\% & \textbf{96\%} & 64\% \\ 
\midrule
Sampled code line correctness & 85\% & 80\% & \textbf{93\%} & 88\%             \\
\bottomrule
\end{tabular}
\end{table}

\begin{table}[t]
\caption{Explanation correctness based on the manual validation}
  \label{tab:correct}
  \centering
\begin{tabular}{@{}cllll@{}}
\toprule
Pair & Groundtruth & GraphCodeBERT & Size-4 & Size-8 \\
\midrule
CP1  & Clone     & Non-clone & Correct & Correct \\
CP2  & Clone     & Non-clone & Correct & Correct \\
CP3  & Clone     & Clone     & Correct & Correct \\
CP4  & Clone     & Non-clone & Correct   & Correct \\
CP5  & Clone     & Clone     & Correct & Correct \\
CP6  & Non-clone & Clone     & Correct & Incorrect  \\
CP7  & Non-clone & Clone     & Correct & Incorrect  \\
CP8  & Non-clone & Clone     & Correct   & Incorrect  \\
CP9  & Non-clone & Non-clone & Correct   & Incorrect  \\
CP10 & Non-clone & Non-clone & Incorrect  & Correct \\ 
\bottomrule
\end{tabular}
\end{table}

\begin{table}[t]
\caption{Classifications of the Explanations}
  \label{tab:classification}
  \centering
\begin{tabular}{lrr}
\toprule
Classification & Size-4 & Size-8 \\
\midrule
Good explanation & 9 & 10 \\
Bad explanation (no example \& irrelevant) & 1 & 0 \\
\bottomrule
\end{tabular}
\end{table}

\subsubsection{RQ1: To what extent can LLMs explain the behavior of code clone detectors?} 
To answer this RQ, we use the results from both the automated validation and the manual validation.

\textbf{Automated validation:~}
The results of the automated validation are shown in Table~\ref{tab:percentage}. The best results are highlighted using bold text. Using the default temperature, our approach with size-4 KLN samples gave 90\% correct explanations. For size-8, our approach gave 78\% correct explanations. Looking deeper into the types of code pairs to be explained, we found that the correctness of the clone pairs is higher at 92\% and 96\% for size-4 and size-8, respectively, compared to 88\% and 60\% of the non-clone pairs. Regarding the accuracy of the sampled code line examples, the approach gave 85\% correct code lines for size-4, and 80\% for size-8.

Interestingly, by reducing the temperature to zero, we observed some improved results. Our approach with size-4 and size-8 examples gave 98\% and 78\% correct explanations, respectively. When looking into the clone pairs and the non-clone pairs, the correctness was also improved in both categories. For clone pairs, our approach gave 100\% and 92\% correct explanations for KLN sample size-4 and size-8, respectively. For non-clone pairs, our approach gave 96\% and 64\% correct explanations for size-4 and size-8, respectively. The correctness of the code line examples was also improved. The approach gave 93\% correct code line examples for size-4, and 88\% for size-8. We can see that the explanations using zero temperature outperform the default temperature in all cases.

\begin{figure}[tb]
    \centering
    \includegraphics[width=\columnwidth]{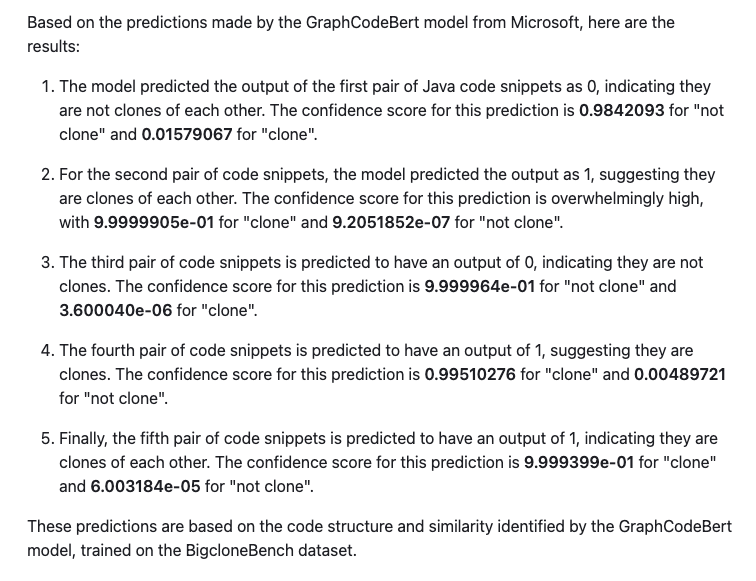}
    \caption{A bad explanation}
    \label{fig:bad_explanation}
\end{figure}

\textbf{Manual validation:~} 
To complement the automated validation, we resort to the qualitative results from the manual validation.
The Cohen's Kappa of the manual validation (95\% confidence interval) for the correctness of explanation is 0.73, which means substantial agreement, and for the explanation quality is 0.38, which indicates a fair agreement. 

As shown in Table \ref{tab:correct}, we found that our technique gave the correct explanations in 15 out of 20 cases. Most of the incorrect explanations are from the non-clone pairs based on the ground truth, but predicted as clone pair by GraphCodeBERT, using the prompts with 8 examples (4 out of the 5 incorrect explanations). This is also similar to the results from the automated validation that the non-clone pairs have lower explanation accuracy. 
We suspect that the higher number of examples may negatively affect the correctness of the explanations, as a longer prompt may confuse the LLM. 
Regarding the quality of the explanation, as shown in Table~\ref{tab:classification}, we found that our approach offers a good explanation in 19 out of 20 cases (95\%). For size 4, there were 9 good explanations, with 7 explanations containing some bad code line examples, and one bad explanation. A bad explanation is shown in Figure~\ref{fig:bad_explanation} (a good explanation is previously shown in  Figure~\ref{fig:explanation}). We can observe that it does not explain the target code pair nor give code line examples, but instead provides the explanation for each of the code pairs sampled from the local neighborhood. For size 8, there was no bad explanation (with 7 explanations containing some bad code line examples).

In addition, we observed that the issue of bad code line examples is one of the challenges, which we aim to investigate in our future work by adjusting the prompt to be more specific (e.g., ``\textit{Show five non-boilerplate lines...}'') or using more sophisticated prompting strategies that guide the LLM to focus on semantic logic rather than syntactic structure. For example, asking it to first summarize the core logic of each snippet and then select lines that correspond to that logic.

\begin{tcolorbox}
Answer to RQ1: LLMs can generate explanations for code clone detection results with up to 98\% accuracy. It performs better when generating explanations for the clone pairs and offers a good explanation 95\% of the time. Moreover, the code line examples are chosen with 93\% accuracy. Nonetheless, the quality of the chosen code line examples needs to be improved.
\end{tcolorbox}

\paragraph{RQ1.1: Given a clone prediction from a blackbox detector, how well does an LLM realize a relevant explanation?} A positive prediction refers to when the model's prediction aligns with the ground truth. As shown in Table~\ref{tab:correct}, the majority of positive predictions (i.e., the pairs CP3, CP5, CP9, and CP10) with 4 and 8 examples receive relevant explanations in 3 out of 4 pairs (75\%).

\begin{tcolorbox}
Answer to RQ1.1: Our approach can be used to explain the clone predictions of clone/non-clone pairs in 3 out of 4 cases.
\end{tcolorbox}

\paragraph{RQ1.2: Given a non-clone prediction from a blackbox detector, how well does an LLM realize a relevant explanation?} A negative prediction is when the model's prediction contrasts with the ground truth. From Table \ref{tab:correct}, we found that for negative predictions (i.e., the pairs CP1, CP2, CP4, CP6, CP7, CP8), the prompts with 4 examples give relevant explanations in 6 out of 6 pairs (100\%), while the prompts with 8 examples give relevant explanation in only 3 out of 6 pairs (50\%). This demonstrates that our approach can also be used to explain the negative predictions of clone/non-clone pairs with some success, with 4 examples providing more correct results than 8 examples.

\begin{tcolorbox}
Answer to RQ1.2: Our approach can be used to explain the non-clone predictions of clone/non-clone pairs with results ranging from 50\% to 100\%. Using 4 KLN samples outperforms using 8 KLN samples. 
\end{tcolorbox}
    
\subsubsection{RQ2: What is the effective number of local neighborhood examples?} 
From the result of automated validation (see Table~\ref{tab:percentage}), the 4 KLN samples give higher explanation accuracy than the 8 KLN samples for both the default temperature and the zero temperature.
Moreover, from the result of manual validation (see Table~\ref{tab:correct}), having 4 neighborhood examples offers more correct explanations compared to 8 examples (9 correct explanations compared to 6 correct explanations). Nonetheless, from Table~\ref{tab:classification}, the quality of the explanations of 8 locally neighbor examples is slightly better than 4 examples (10 compared to 9).

Although a previous study found that 32 examples offer the best performance~\cite{Gao2023}, we found that prompts with much lower example sizes of 4 outperform the prompts with 8 examples in terms of explanation of clone detection prediction. 
Lastly, many of the code line examples are basic statements such as variable initialization or loop headers rather than statements that capture the code semantics (e.g., array value swapping). 

\begin{tcolorbox}
Answer to RQ2: Using the KLN samples of size 4 gives better explanation accuracy than its size-8 counterpart. Nonetheless, the quality of the explanation of size 8 is better.
\end{tcolorbox}

\subsubsection{RQ3: What are the effects of the LLM temperature on generating the explanations?} 
As shown in Table~\ref{tab:percentage} and discussed in the answer to RQ1, we found from the automated validation that adjusting the temperature of the LLM to zero offers higher explanation accuracy, compared to the default temperature. 
The highest explanation accuracy is by using KLN size-4 with zero temperature.
This is potentially due to the reduction of the LLM hallucination under low temperature. Thus, the generated explanation and the sampled code lines are correctly drawn from the original code snippet. Unfortunately, we did not try generating the explanations using the zero temperature for our manual analysis. So, we could not compare the quality of the explanations based on the temperature. We leave that as our future work. 

\begin{tcolorbox}
Answer to RQ3: Lowering the LLM's temperature to zero increases the accuracy of the explanation.
\end{tcolorbox}

\subsubsection{RQ4: How are the sampled code lines selected for the explanations?}

\begin{figure}
    \centering
    \includegraphics[width=\columnwidth]{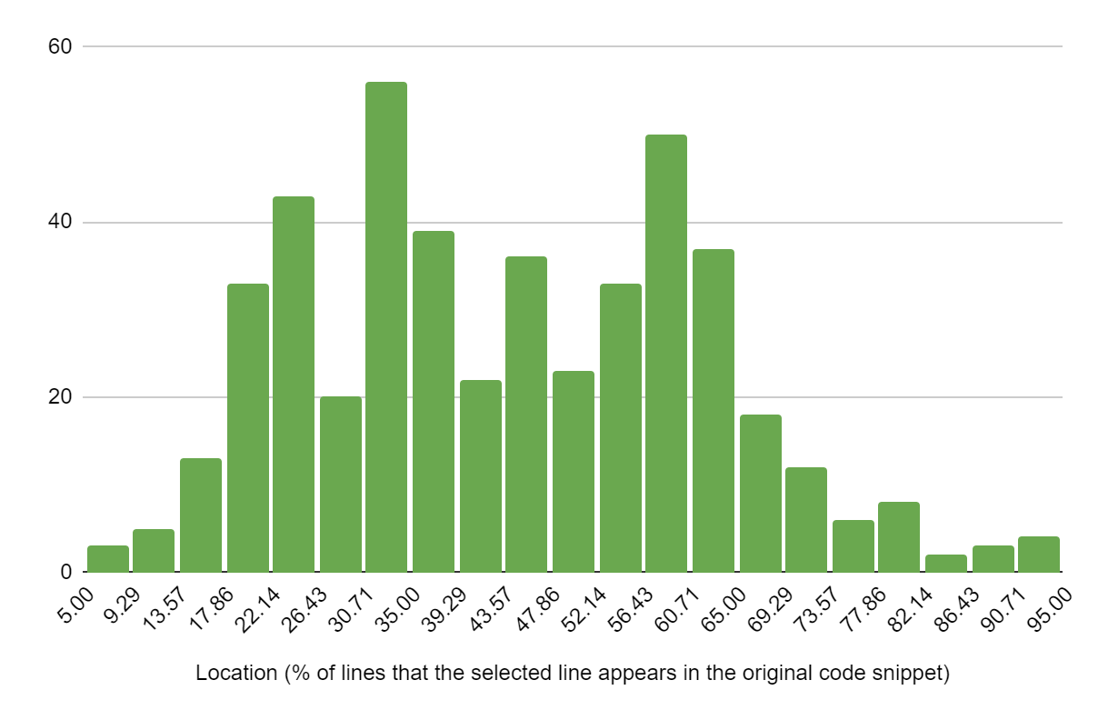}
    \caption{Distribution of the locations of selected code lines (size-4, temperatures=0)}
    \label{fig:codeline_dist}
\end{figure}

To answer this RQ, we looked at the locations of the code line examples in the original code snippet pair. We picked the setting with the highest explanation accuracy for this analysis, i.e., size-4 with zero temperature (see Table~\ref{tab:percentage}). Figure~\ref{fig:codeline_dist} shows the distribution of the 2,000 code line examples (10 code line examples for each of the 200 analyzed code pairs) in terms of the distance from the first line (percentage). For example, if a code line example appears on the first line of a code snippet with 100 lines, we mark that it appears at the 1\% distance. If it appears on line 50, we mark it at 50\% distance. If it appears on the last line, it has 100\% distance.

From the figure, we can observe that the code line examples were chosen across the original code. Nonetheless, the majority of the code line examples are selected starting from a distance of 17.86\% to 65\% from the beginning of the file. 
The frequencies drop dramatically at the beginning and at the end of the file.

\begin{tcolorbox}
Answer to RQ4: The selected code lines are mostly from the middle part of the original code, containing the main logic of the code.
\end{tcolorbox}

\subsubsection{Additional Observations}
We summarize the unexpected or interesting explanations we found when investigating the LLM responses here. 

First, in some of the explanations, the code line examples may not be the same as the original code, despite their resemblance (i.e., good explanation but bad code line examples). This may potentially be because of the hallucination.
More prompting strategies may be needed to mitigate these issues.
Second, some explanations generated by the LLM  contained more than 5 lines of code or did not give any code. Third, some of the explanations of the code line examples do not match the code lines themselves. Fourth, many of the code line examples focused on basic statements such as variable initialization, loop header, or output printing rather than the code that captures the semantics of the code (e.g., swapping values in an array). More prompting strategies are needed to mitigate these issues.

\section{Related Work}

\subsection{Machine-learning-based Code Clone Detection}
Machine learning (ML) techniques have demonstrated promising results in code similarity measurement and clone detection, owing to their capacity to learn and recognize complex patterns. These approaches typically aim to classify or cluster similar code fragments by training on datasets of known similar and dissimilar code~\cite{ZakeriNasrabadi2023}. 
Some of the ML-based clone detection models are classification models that are explainable, like Random Forest, have been used to identify clones at the package level~\cite{Cesare2013}. Clustering algorithms such as DBSCAN~\cite{Joshi2015} and k-means~\cite{Keivanloo2015} have also been employed for detecting clones at the function level.

With the advancement of machine learning techniques, more complex ML models, i.e., deep learning models, are increasingly adopted in clone detection studies. Neural network models including Artificial Neural Networks (ANN)~\cite{Mostaeen2019,Mostaeen2020,Li2017}, Siamese neural networks (SNN) and Siamese deep neural networks (SDNN)~\cite{Patel2022,Saini2018}, Code embedding approaches like code2vec~\cite{Alon2019a} and code2seq~\cite{Alon2019b}, are used in the recent clone studies. 
Pre-trained models such as CodeBERT~\cite{Feng2020}, Moreover, GraphCodeBERT~\cite{Guo2021}, and UnixCoder~\cite{Guo2022} have been adapted for code clone detection, including cross-language code clone detection.
Nonetheless, providing explanations for these deep learning models is still a challenging task as they operate as a blackbox model.

\subsection{Explainable AI for Software Engineering}
Explainable AI for Software Engineering (XAI4SE) is a pressing concern for the software industry and academia due to the widespread adoption of AI/ML tools, such as code completion and defect prediction, which are often based on complex, black-box deep learning models~\cite{Tantithamthavorn2023}. 
Practitioners often distrust the outputs of the AI-based techniques because they do not understand the underlying rationale for predictions~\cite{xai4sebook,Dam2018}.
XAI4SE has been adopted in many stages in the software development life cycle~\cite{Mohammadkhani2023}. 
The most prominent stage is software maintenance, such as defect prediction, to specifically find buggy lines or tokens within code, enhancing the practicality of defect prediction~\cite{Jiarpakdee2022,Khanan2020,Pornprasit2021}, code reviews~\cite{Ochodek2020}, technical debt detection~\cite{Ren2019}, and valid bug report detection~\cite{He2020}.
It is also used during software development (e.g., method name prediction and variable misuse detection~\cite{Rabin2021}), software management (e.g., effort prediction~\cite{Basgalupp2013}), and software requirements (e.g., requirements engineering and requirements quality~\cite{Moreno2020}).

\subsubsection*{LLMs for Generating Explanations}
The study by Valenzuela~\cite{Valenzuela2025} addresses the significant challenge developers face in diagnosing GitHub Actions (GA) workflow failures due to the volume, complexity, and unstructured nature of error logs using LLMs. They conducted a mixed-methods feasibility study involving 31 developers, evaluating LLM-generated explanations of the GA error logs and found that over 80\% of developers rated the explanations positively for correctness and clarity in simpler/smaller logs. The study suggests that the explanations should be adapted to user expertise, as less experienced developers preferred detailed explanations while seasoned developers favored concise summaries.

\section{Threats to Validity}
\paragraph*{Internal validity}The samples of the 5 clone pairs and 5 non-clone pairs may not give a representative sample. They may also cause some biases in generating the explanations due to the different difficulties of each question in the Google Code Jam competition. We mitigated this threat by random sampling across different questions. Moreover, the nondeterminism of LLMs may cause the results to be different in every response. We mitigated it by generating the explanations five times for each prompt in the automated validation. Lastly, the manual validation may be subject to some human biases and errors. We mitigated this threat by employing two validators for each explanation and recruited a third validator if needed.
\paragraph*{External validity}
The study is based only on the Java programming language, and the findings may not be generalized to other programming languages. The data used in the study is based on the Google Code Jam programming competition, which offers programs with perform the same task with different implementations, and may not be generalized to other types of code clones. Lastly, the explanations were generated by one LLM model, OpenAI GPT-4. Thus, the findings may not be generalized to other LLMs. 

\section{Conclusion}
We present a novel approach using in-context learning of LLMs to explain the predictions of GraphCodeBERT, an ML-based clone detector. The experimental results show that our approach can be used to provide explanations for clone and non-clone pairs with moderate success. It can generate explanations for code clone detection results with up to 98\% accuracy and offer good explanations 95\% of the time. Moreover, we discovered that using the local neighborhood of size 4 offers more accurate results than size 8, but with lower explanation quality. Lastly, we found that zero temperature generates a higher accuracy than the default temperature.

We plan to expand the work by (1) strengthening the findings in the manual validation by running the same prompt multiple times to mitigate the randomness in the LLM responses and also cover the zero temperature, (2) using a predefined distribution on the neighborhoods and investigating their effects, (3) varying the temperature of the LLM, (4) comparing the explanations with and without giving the question text from GCJ questions, and (5) employing some distance measure (e.g., cosine distance, Euclidean distance) with other code representations (e.g., trees, graphs, embeddings) as a measure of proximity for the local neighborhood instead of relying on the GCJ questions. 

We believe that using LLMs with the concept of KLN as a post hoc explainer for code clone detection and for other software engineering tasks is promising and still much underexplored. We aim to dive deeper into studying this direction in the future.

The replication package of our study is available at \url{https://doi.org/10.5281/zenodo.11652366}.

\bibliographystyle{IEEEtran}
\bibliography{references}

% Generated by IEEEtran.bst, version: 1.14 (2015/08/26)
\begin{thebibliography}{10}
\providecommand{\url}[1]{#1}
\csname url@samestyle\endcsname
\providecommand{\newblock}{\relax}
\providecommand{\bibinfo}[2]{#2}
\providecommand{\BIBentrySTDinterwordspacing}{\spaceskip=0pt\relax}
\providecommand{\BIBentryALTinterwordstretchfactor}{4}
\providecommand{\BIBentryALTinterwordspacing}{\spaceskip=\fontdimen2\font plus
\BIBentryALTinterwordstretchfactor\fontdimen3\font minus \fontdimen4\font\relax}
\providecommand{\BIBforeignlanguage}[2]{{%
\expandafter\ifx\csname l@#1\endcsname\relax
\typeout{** WARNING: IEEEtran.bst: No hyphenation pattern has been}%
\typeout{** loaded for the language `#1'. Using the pattern for}%
\typeout{** the default language instead.}%
\else
\language=\csname l@#1\endcsname
\fi
#2}}
\providecommand{\BIBdecl}{\relax}
\BIBdecl

\bibitem{Rattan2013}
``{Software clone detection: A systematic review},'' \emph{Information and Software Technology}, vol.~55, no.~7, pp. 1165--1199, Jul 2013.

\bibitem{Roy2009}
C.~K. Roy, J.~R. Cordy, and R.~Koschke, ``{Comparison and evaluation of code clone detection techniques and tools: A qualitative approach},'' \emph{Science of Computer Programming}, vol.~74, no.~7, pp. 470--495, 2009.

\bibitem{Kapser2008}
C.~J. Kapser and M.~W. Godfrey, ``{Cloning considered harmful considered harmful: patterns of cloning in software},'' \emph{Empirical Software Engineering}, vol.~13, no.~6, pp. 645--692, Dec 2008.

\bibitem{Krinke2008}
J.~Krinke, ``{Is Cloned Code More Stable than Non-cloned Code?}'' in \emph{Proceedings of 8th IEEE International Working Conference on Source Code Analysis and Manipulation (SCAM '08)}, 2008, pp. 57--66.

\bibitem{Balazinska2000}
M.~Balazinska, E.~Merlo, M.~Dagenais, B.~Lague, and K.~Kontogiannis, ``Advanced clone-analysis to support object-oriented system refactoring,'' in \emph{Proceedings of 7th Working Conference on Reverse Engineering (RE)}.\hskip 1em plus 0.5em minus 0.4em\relax IEEE, 2000, pp. 98--107.

\bibitem{Luan2019}
S.~Luan, D.~Yang, C.~Barnaby, K.~Sen, and S.~Chandra, ``{Aroma}: code recommendation via structural code search,'' \emph{Proceedings of the ACM on Programming Languages}, vol.~3, no. OOPSLA, pp. 1--28, Oct 2019.

\bibitem{Prechelt2002}
L.~Prechelt, G.~Malpohl, and M.~Philippsen, ``Finding plagiarisms among a set of programs with {JPlag},'' \emph{Journal Of Universal Computer Science}, vol.~8, no.~11, pp. 1016--1038, 2002.

\bibitem{Davies2013}
J.~Davies, D.~M. German, M.~W. Godfrey, and A.~Hindle, ``{Software Bertillonage},'' \emph{Empirical Software Engineering}, vol.~18, no.~6, pp. 1195--1237, Dec 2013.

\bibitem{Farhadi2015}
M.~R. Farhadi, B.~C. Fung, Y.~B. Fung, P.~Charland, S.~Preda, and M.~Debbabi, ``{Scalable code clone search for malware analysis},'' \emph{Digital Investigation}, vol.~15, pp. 46--60, Dec 2015.

\bibitem{Saini2018}
V.~Saini, F.~Farmahinifarahani, Y.~Lu, P.~Baldi, and C.~V. Lopes, ``{Oreo}: detection of clones in the twilight zone,'' in \emph{Proceedings of the 2018 26th ACM Joint Meeting on European Software Engineering Conference and Symposium on the Foundations of Software Engineering (ESEC/FSE '18)}, jun 2018, pp. 354--365.

\bibitem{Roy2008}
C.~K. Roy and J.~Cordy, ``{NICAD}: Accurate detection of near-miss intentional clones using flexible pretty-printing and code normalization,'' in \emph{2008 16th IEEE International Conference on Program Comprehension}.\hskip 1em plus 0.5em minus 0.4em\relax IEEE, jun 2008, pp. 172--181.

\bibitem{Sajnani2016}
H.~Sajnani, V.~Saini, J.~Svajlenko, C.~K. Roy, and C.~V. Lopes, ``{SourcererCC}: Scaling code clone detection to big-code,'' in \emph{Proceedings of the 38th International Conference on Software Engineering (ICSE '16)}.\hskip 1em plus 0.5em minus 0.4em\relax ACM Press, 2016, pp. 1157--1168.

\bibitem{Ragkhitwetsagul2019}
C.~Ragkhitwetsagul and J.~Krinke, ``{Siamese}: scalable and incremental code clone search via multiple code representations,'' \emph{Empirical Software Engineering}, vol.~24, no.~4, pp. 2236--2284, aug 2019.

\bibitem{Jiang2007a}
L.~Jiang, G.~Misherghi, Z.~Su, and S.~Glondu, ``{DECKARD}: Scalable and accurate tree-based detection of code clones,'' in \emph{Proceedings of the 29th International Conference on Software Engineering (ICSE'07)}, 2007, pp. 96--105.

\bibitem{Krinke2001}
J.~Krinke, ``Identifying similar code with program dependence graphs,'' in \emph{Proceedings of 8th Working Conference on Reverse Engineering (WCRE '08)}.\hskip 1em plus 0.5em minus 0.4em\relax IEEE Comput. Soc, 2001, pp. 301--309.

\bibitem{Zhang2019}
J.~Zhang, X.~Wang, H.~Zhang, H.~Sun, K.~Wang, and X.~Liu, ``{A Novel Neural Source Code Representation Based on Abstract Syntax Tree},'' in \emph{Proceedings of the IEEE/ACM 41st International Conference on Software Engineering (ICSE '19)}, vol. 2019-May, 2019, pp. 783--794.

\bibitem{lei2022deep}
M.~Lei, H.~Li, J.~Li, N.~Aundhkar, and D.-K. Kim, ``Deep learning application on code clone detection: A review of current knowledge,'' \emph{Journal of Systems and Software}, vol. 184, p. 111141, 2022.

\bibitem{mostaeen2020machine}
G.~Mostaeen, B.~Roy, C.~K. Roy, K.~Schneider, and J.~Svajlenko, ``A machine learning based framework for code clone validation,'' \emph{Journal of Systems and Software}, vol. 169, p. 110686, 2020.

\bibitem{wang2020detecting}
W.~Wang, G.~Li, B.~Ma, X.~Xia, and Z.~Jin, ``Detecting code clones with graph neural network and flow-augmented abstract syntax tree,'' in \emph{2020 IEEE 27th International Conference on Software Analysis, Evolution and Reengineering (SANER)}.\hskip 1em plus 0.5em minus 0.4em\relax IEEE, 2020, pp. 261--271.

\bibitem{Guo2021}
D.~Guo, S.~Ren, S.~Lu, Z.~Feng, D.~Tang, S.~Liu, L.~Zhou, N.~Duan, A.~Svyatkovskiy, S.~Fu, M.~Tufano, S.~K. Deng, C.~Clement, D.~Drain, N.~Sundaresan, J.~Yin, D.~Jiang, and M.~Zhou, ``{GraphCodeBERT: Pre-training Code Representations with Data Flow},'' in \emph{Proceedings of the 9th International Conference on Learning Representations (ICLR '21)}, Sep 2021, pp. 1--18.

\bibitem{sajnani2016sourcerercc}
H.~Sajnani, V.~Saini, J.~Svajlenko, C.~K. Roy, and C.~V. Lopes, ``Sourcerercc: Scaling code clone detection to big-code,'' in \emph{Proceedings of the 38th international conference on software engineering}, 2016, pp. 1157--1168.

\bibitem{li2021secnn}
Z.~Li, Y.~Wu, B.~Peng, X.~Chen, Z.~Sun, Y.~Liu, and D.~Yu, ``Secnn: A semantic cnn parser for code comment generation,'' \emph{Journal of Systems and Software}, vol. 181, p. 111036, 2021.

\bibitem{Feng2020}
Z.~Feng, D.~Guo, D.~Tang, N.~Duan, X.~Feng, M.~Gong, L.~Shou, B.~Qin, T.~Liu, D.~Jiang, and M.~Zhou, ``{CodeBERT: A Pre-Trained Model for Programming and Natural Languages},'' in \emph{Proceedings of the Findings of the Association for Computational Linguistics (EMNLP '20)}, 2020, pp. 1536--1547.

\bibitem{xai4sebook}
\BIBentryALTinterwordspacing
C.~Tantithamthavorn and J.~Jiarpakdee, \emph{Explainable AI for Software Engineering}.\hskip 1em plus 0.5em minus 0.4em\relax Monash University, 2021, retrieved 2021-05-17. [Online]. Available: \url{http://xai4se.github.io/}
\BIBentrySTDinterwordspacing

\bibitem{ribeiro2016should}
M.~T. Ribeiro, S.~Singh, and C.~Guestrin, ``" why should i trust you?" explaining the predictions of any classifier,'' in \emph{Proceedings of the 22nd ACM SIGKDD international conference on knowledge discovery and data mining}, 2016, pp. 1135--1144.

\bibitem{palacio2021xai}
S.~Palacio, A.~Lucieri, M.~Munir, S.~Ahmed, J.~Hees, and A.~Dengel, ``Xai handbook: towards a unified framework for explainable ai,'' in \emph{Proceedings of the IEEE/CVF International Conference on Computer Vision}, 2021, pp. 3766--3775.

\bibitem{chou2022counterfactuals}
Y.-L. Chou, C.~Moreira, P.~Bruza, C.~Ouyang, and J.~Jorge, ``Counterfactuals and causability in explainable artificial intelligence: Theory, algorithms, and applications,'' \emph{Information Fusion}, vol.~81, pp. 59--83, 2022.

\bibitem{mann2020language}
B.~Mann, N.~Ryder, M.~Subbiah, J.~Kaplan, P.~Dhariwal, A.~Neelakantan, P.~Shyam, G.~Sastry, A.~Askell, S.~Agarwal \emph{et~al.}, ``Language models are few-shot learners,'' \emph{arXiv preprint arXiv:2005.14165}, 2020.

\bibitem{miller2017explainable}
T.~Miller, P.~Howe, and L.~Sonenberg, ``Explainable ai: Beware of inmates running the asylum or: How i learnt to stop worrying and love the social and behavioural sciences,'' 2017.

\bibitem{shap}
S.~Lundberg and S.-I. Lee, ``A unified approach to interpreting model predictions,'' 2017.

\bibitem{Lime}
M.~T. Ribeiro, S.~Singh, and C.~Guestrin, ``"why should i trust you?": Explaining the predictions of any classifier,'' 2016.

\bibitem{pawar2024impact}
U.~Pawar, C.~Beder, O.~Ruairi, O.~Donna \emph{et~al.}, ``On the impact of neighbourhood sampling to satisfy sufficiency and necessity criteria in explainable ai,'' in \emph{Causal Learning and Reasoning}.\hskip 1em plus 0.5em minus 0.4em\relax PMLR, 2024, pp. 570--586.

\bibitem{brown2020language}
T.~Brown, B.~Mann, N.~Ryder, M.~Subbiah, J.~D. Kaplan, P.~Dhariwal, A.~Neelakantan, P.~Shyam, G.~Sastry, A.~Askell \emph{et~al.}, ``Language models are few-shot learners,'' \emph{Advances in neural information processing systems}, vol.~33, pp. 1877--1901, 2020.

\bibitem{wei2022chain}
J.~Wei, X.~Wang, D.~Schuurmans, M.~Bosma, F.~Xia, E.~Chi, Q.~V. Le, D.~Zhou \emph{et~al.}, ``Chain-of-thought prompting elicits reasoning in large language models,'' \emph{Advances in neural information processing systems}, vol.~35, pp. 24\,824--24\,837, 2022.

\bibitem{GPT}
\BIBentryALTinterwordspacing
OpenAI, ``Introducing chatgpt,'' (accessed: 07.06.2024). [Online]. Available: \url{https://openai.com/index/chatgpt/}
\BIBentrySTDinterwordspacing

\bibitem{Claude-2}
\BIBentryALTinterwordspacing
Anthropic, ``Meet claude,'' (accessed: 07.06.2024). [Online]. Available: \url{https://www.anthropic.com/claude}
\BIBentrySTDinterwordspacing

\bibitem{Gemini}
\BIBentryALTinterwordspacing
G.~DeepMind, ``Gemini models,'' (accessed: 07.06.2024). [Online]. Available: \url{https://deepmind.google/technologies/gemini/}
\BIBentrySTDinterwordspacing

\bibitem{Wei2022ChainOT}
\BIBentryALTinterwordspacing
J.~Wei, X.~Wang, D.~Schuurmans, M.~Bosma, E.~H. hsin Chi, F.~Xia, Q.~Le, and D.~Zhou, ``Chain of thought prompting elicits reasoning in large language models,'' \emph{ArXiv}, vol. abs/2201.11903, 2022. [Online]. Available: \url{https://api.semanticscholar.org/CorpusID:246411621}
\BIBentrySTDinterwordspacing

\bibitem{Hendy2023HowGA}
\BIBentryALTinterwordspacing
A.~Hendy, M.~G. Abdelrehim, A.~Sharaf, V.~Raunak, M.~Gabr, H.~Matsushita, Y.~J. Kim, M.~Afify, and H.~H. Awadalla, ``How good are gpt models at machine translation? a comprehensive evaluation,'' \emph{ArXiv}, vol. abs/2302.09210, 2023. [Online]. Available: \url{https://api.semanticscholar.org/CorpusID:257038384}
\BIBentrySTDinterwordspacing

\bibitem{Bubeck2023SparksOA}
S.~Bubeck, V.~Chandrasekaran, R.~Eldan, J.~A. Gehrke, E.~Horvitz, E.~Kamar, P.~Lee, Y.~T. Lee, Y.-F. Li, S.~M. Lundberg, H.~Nori, H.~Palangi, M.~T. Ribeiro, and Y.~Zhang, ``Sparks of artificial general intelligence: Early experiments with gpt-4,'' \emph{ArXiv}, vol. abs/2303.12712, 2023.

\bibitem{kroeger2024large}
N.~Kroeger, D.~Ley, S.~Krishna, C.~Agarwal, and H.~Lakkaraju, ``Are large language models post hoc explainers?'' 2024.

\bibitem{vu2020progressive}
T.-K. Vu, T.~Racharak, S.~Tojo, H.-T. Nguyen, and L.~M.~N. 0001, ``Progressive training in recurrent neural networks for chord progression modeling.'' in \emph{ICAART (2)}, 2020, pp. 89--98.

\bibitem{smilkov2017smoothgrad}
D.~Smilkov, N.~Thorat, B.~Kim, F.~Viégas, and M.~Wattenberg, ``Smoothgrad: removing noise by adding noise,'' 2017.

\bibitem{liu2023lost}
N.~F. Liu, K.~Lin, J.~Hewitt, A.~Paranjape, M.~Bevilacqua, F.~Petroni, and P.~Liang, ``Lost in the middle: How language models use long contexts,'' 2023.

\bibitem{Svajlenko2022}
J.~Svajlenko and C.~K. Roy, ``{BigCloneBench}: A retrospective and roadmap,'' in \emph{2022 IEEE 16th International Workshop on Software Clones (IWSC)}, oct 2022, pp. 8--9.

\bibitem{Zhao2018}
G.~Zhao and J.~Huang, ``{DeepSim: deep learning code functional similarity},'' in \emph{Proceedings of the 2018 26th ACM Joint Meeting on European Software Engineering Conference and Symposium on the Foundations of Software Engineering}.\hskip 1em plus 0.5em minus 0.4em\relax ACM, 2018, pp. 141--151.

\bibitem{Krinke2021}
J.~Krinke and C.~Ragkhitwetsagul, ``{Code Similarity in Clone Detection},'' in \emph{Code Clone Analysis}, K.~Inoue and C.~K. Roy, Eds.\hskip 1em plus 0.5em minus 0.4em\relax Springer Singapore, 2021, pp. 135--160.

\bibitem{Choi2023}
E.~Choi, N.~Fuke, Y.~Fujiwara, N.~Yoshida, and K.~Inoue, ``{Investigating the Generalizability of Deep Learning-based Clone Detectors},'' in \emph{2023 IEEE/ACM 31st International Conference on Program Comprehension (ICPC)}, vol.~3, 2023, pp. 181--185.

\bibitem{Su2016a}
F.-H. Su, J.~Bell, G.~Kaiser, and S.~Sethumadhavan, ``{Identifying functionally similar code in complex codebases},'' in \emph{Proceedings of the IEEE 24th International Conference on Program Comprehension (ICPC '16)}.\hskip 1em plus 0.5em minus 0.4em\relax IEEE, 2016, pp. 1--10.

\bibitem{Su2016b}
F.-H. Su, J.~Bell, K.~Harvey, S.~Sethumadhavan, G.~Kaiser, and T.~Jebara, ``{Code relatives: detecting similarly behaving software},'' in \emph{Proceedings of the 2016 24th ACM SIGSOFT International Symposium on Foundations of Software Engineering}.\hskip 1em plus 0.5em minus 0.4em\relax ACM, 2016, pp. 702--714.

\bibitem{Gao2023}
S.~Gao, X.-C. Wen, C.~Gao, W.~Wang, H.~Zhang, and M.~R. Lyu, ``{What Makes Good In-Context Demonstrations for Code Intelligence Tasks with LLMs?}'' in \emph{2023 38th IEEE/ACM International Conference on Automated Software Engineering (ASE)}.\hskip 1em plus 0.5em minus 0.4em\relax IEEE, 2023, pp. 761--773.

\bibitem{ZakeriNasrabadi2023}
M.~Zakeri-Nasrabadi, S.~Parsa, M.~Ramezani, C.~Roy, and M.~Ekhtiarzadeh, ``A systematic literature review on source code similarity measurement and clone detection: Techniques, applications, and challenges,'' \emph{Journal of Systems and Software}, vol. 204, p. 111796, 2023.

\bibitem{Cesare2013}
S.~Cesare, Y.~Xiang, and J.~Zhang, ``Clonewise -- detecting package-level clones using machine learning,'' in \emph{Security and Privacy in Communication Networks}.\hskip 1em plus 0.5em minus 0.4em\relax Springer International Publishing, 2013, pp. 197--215.

\bibitem{Joshi2015}
B.~Joshi, P.~Budhathoki, W.~L. Woon, and D.~Svetinovic, ``Software clone detection using clustering approach,'' in \emph{Proceeings, Part II, of the 22nd International Conference on Neural Information Processing - Volume 9490 (ICONIP '15)}, 2015, p. 520–527.

\bibitem{Keivanloo2015}
I.~Keivanloo, F.~Zhang, and Y.~Zou, ``Threshold-free code clone detection for a large-scale heterogeneous java repository,'' in \emph{2015 IEEE 22nd International Conference on Software Analysis, Evolution, and Reengineering (SANER '15)}, 2015, pp. 201--210.

\bibitem{Mostaeen2019}
G.~Mostaeen, J.~Svajlenko, B.~Roy, C.~K. Roy, and K.~A. Schneider, ``Clonecognition: machine learning based code clone validation tool,'' in \emph{Proceedings of the 2019 27th ACM Joint Meeting on European Software Engineering Conference and Symposium on the Foundations of Software Engineering}, ser. ESEC/FSE 2019, 2019, p. 1105–1109.

\bibitem{Mostaeen2020}
G.~Mostaeen, B.~Roy, C.~K. Roy, K.~Schneider, and J.~Svajlenko, ``A machine learning based framework for code clone validation,'' \emph{Journal of Systems and Software}, vol. 169, p. 110686, 2020.

\bibitem{Li2017}
L.~Li, H.~Feng, W.~Zhuang, N.~Meng, and B.~Ryder, ``Cclearner: A deep learning-based clone detection approach,'' in \emph{Proceedings of the 2017 IEEE International Conference on Software Maintenance and Evolution (ICSME '17)}, 2017, pp. 249--260.

\bibitem{Patel2022}
S.~Patel and R.~Sinha, ``Combining holistic source code representation with siamese neural networks for detecting code clones,'' in \emph{Testing Software and Systems}, 2022, pp. 148--159.

\bibitem{Alon2019a}
U.~Alon, M.~Zilberstein, O.~Levy, and E.~Yahav, ``code2vec: learning distributed representations of code,'' \emph{Proceedings of the ACM on Programming Languages}, vol.~3, no. POPL, Jan. 2019.

\bibitem{Alon2019b}
\BIBentryALTinterwordspacing
U.~Alon, S.~Brody, O.~Levy, and E.~Yahav, ``code2seq: Generating sequences from structured representations of code,'' 2019. [Online]. Available: \url{https://arxiv.org/abs/1808.01400}
\BIBentrySTDinterwordspacing

\bibitem{Guo2022}
D.~Guo, S.~Lu, N.~Duan, Y.~Wang, M.~Zhou, and J.~Yin, ``Unixcoder: Unified cross-modal pre-training for code representation,'' \emph{arXiv}, 2022.

\bibitem{Tantithamthavorn2023}
C.~Tantithamthavorn, J.~Cito, H.~Hemmati, and S.~Chandra, ``Explainable ai for se: Challenges and future directions,'' \emph{IEEE Software}, vol.~40, no.~3, pp. 29--33, 2023.

\bibitem{Dam2018}
H.~K. Dam, T.~Tran, and A.~Ghose, ``Explainable software analytics,'' in \emph{Proceedings of the 40th International Conference on Software Engineering: New Ideas and Emerging Results (ICSE NIER '18)}, 2018, p. 53–56.

\bibitem{Mohammadkhani2023}
\BIBentryALTinterwordspacing
A.~H. Mohammadkhani, N.~S. Bommi, M.~Daboussi, O.~Sabnis, C.~Tantithamthavorn, and H.~Hemmati, ``A systematic literature review of explainable ai for software engineering,'' 2023. [Online]. Available: \url{https://arxiv.org/abs/2302.06065}
\BIBentrySTDinterwordspacing

\bibitem{Jiarpakdee2022}
J.~Jiarpakdee, C.~K. Tantithamthavorn, H.~K. Dam, and J.~Grundy, ``{ An Empirical Study of Model-Agnostic Techniques for Defect Prediction Models },'' \emph{IEEE Transactions on Software Engineering}, vol.~48, no.~01, pp. 166--185, Jan. 2022.

\bibitem{Khanan2020}
C.~Khanan, W.~Luewichana, K.~Pruktharathikoon, J.~Jiarpakdee, C.~Tantithamthavorn, M.~Choetkiertikul, C.~Ragkhitwetsagul, and T.~Sunetnanta, ``Jitbot: an explainable just-in-time defect prediction bot,'' in \emph{Proceedings of the 35th IEEE/ACM International Conference on Automated Software Engineering (ASE '20)}, 2021, p. 1336–1339.

\bibitem{Pornprasit2021}
C.~Pornprasit, C.~Tantithamthavorn, J.~Jiarpakdee, M.~Fu, and P.~Thongtanunam, ``Pyexplainer: explaining the predictions of just-in-time defect models,'' in \emph{Proceedings of the 36th IEEE/ACM International Conference on Automated Software Engineering (ASE '21)}, 2022, p. 407–418.

\bibitem{Ochodek2020}
M.~Ochodek, R.~Hebig, W.~Meding, G.~Frost, and M.~Staron, ``Recognizing lines of code violating company-specific coding guidelines using machine learning,'' \emph{Empirical Software Engineering}, vol.~25, no.~1, pp. 220--265, jan 2020.

\bibitem{Ren2019}
X.~Ren, Z.~Xing, X.~Xia, D.~Lo, X.~Wang, and J.~Grundy, ``Neural network-based detection of self-admitted technical debt: From performance to explainability,'' vol.~28, no.~3, Jul. 2019.

\bibitem{He2020}
J.~He, L.~Xu, Y.~Fan, Z.~Xu, M.~Yan, and Y.~Lei, ``Deep learning based valid bug reports determination and explanation,'' in \emph{Proceedings of IEEE 31st International Symposium on Software Reliability Engineering (ISSRE '20)}, 2020, pp. 184--194.

\bibitem{Rabin2021}
M.~R.~I. Rabin, V.~J. Hellendoorn, and M.~A. Alipour, ``Understanding neural code intelligence through program simplification,'' in \emph{Proceedings of the 29th ACM Joint Meeting on European Software Engineering Conference and Symposium on the Foundations of Software Engineering (ESEC/FSE '21)}, 2021, p. 441–452.

\bibitem{Basgalupp2013}
M.~P. Basgalupp, R.~C. Barros, T.~S. da~Silva, and A.~C. P. L.~F. de~Carvalho, ``Software effort prediction: a hyper-heuristic decision-tree based approach,'' in \emph{Proceedings of the 28th Annual ACM Symposium on Applied Computing (SAC '13)}, 2013, p. 1109–1116.

\bibitem{Moreno2020}
V.~Moreno, G.~Génova, E.~Parra, and A.~Fraga, ``Application of machine learning techniques to the flexible assessment and improvement of requirements quality,'' \emph{Software Quality Journal}, vol.~28, no.~4, pp. 1645--1674, dec 2020.

\bibitem{Valenzuela2025}
P.~Valenzuela-Toledo, C.~Wu, S.~Hernandez, A.~Boll, R.~Machacek, S.~Panichella, and T.~Kehrer, ``{Explaining GitHub Actions Failures with Large Language Models: Challenges, Insights, and Limitations},'' in \emph{ICPC'25}, 2025.

\end{thebibliography}
\end{document}